\documentclass{article}
\usepackage[utf8]{inputenc}
\usepackage[margin=1in,top=15 mm]{geometry}
\usepackage{authblk}
\usepackage{setspace}
\usepackage[colorlinks,citecolor=blue,urlcolor=blue,bookmarks=false,hypertexnames=true]{hyperref} 
\usepackage{graphicx}
\usepackage{multirow}
\usepackage{csquotes}
\usepackage[numbers,sort&compress]{natbib}
\bibpunct{[}{]}{,}{n}{}{,}
\usepackage{verbatim}
\usepackage{float}
\usepackage{xcolor}
\usepackage{hyperref}
\usepackage[colorlinks,citecolor=blue,urlcolor=blue,bookmarks=false,hypertexnames=true]{hyperref}
\usepackage[nameinlink,noabbrev]{cleveref}
\usepackage{algorithm2e}
\begin{document}

\sloppy

\title{\textbf{Implicit Geometric Descriptor-Enabled ANN Framework for a Unified Structure-Property Relationship in Architected Nanofibrous Materials}}

\def\correspondingauthor{\footnote{Corresponding author: thevamaran@wisc.edu}}

\author{Bhanugoban Maheswaran}
\author{Komal Chawla}
\author{Abhishek Gupta}

\author{Ramathasan Thevamaran \correspondingauthor{}}

\affil{Department of Mechanical Engineering, University of Wisconsin-Madison, Madison, WI, 53706, USA}


\maketitle

\vspace*{-\baselineskip}
\vspace*{-\baselineskip}
\par\noindent\rule{\textwidth}{0.5pt}
\smallskip


\section*{Abstract}

\doublespacing

Hierarchically architected nanofibrous materials, such as the vertically aligned carbon nanotube (VACNT) foams, draw their exceptional mechanical properties from the interplay of nanoscale size effects and inter-nanotube interactions within and across architectures. However, the distinct effects of these mechanisms, amplified by the architecture, on different mechanical properties remain elusive, limiting their independent tunability for targeted property combinations. Reliance on architecture-specific explicit design parameters further inhibits the development of a unified structure-property relationship rooted in those nanoscale mechanisms. Here, we introduce two implicit geometric descriptors---multi-component shape invariants (MCSI)---in an artificial neural network (ANN) framework to establish a unified structure-property relationship that governs diverse architectures. The MCSIs effectively capture the key nanoscale mechanisms that give rise to the bulk mechanical properties such as specific-energy absorption, peak stress, and average modulus. Exploiting their ability to predict mechanical properties for designs that are even outside of the training data, we propose generalized design strategies to achieve desired mechanical property combinations in architected VACNT foams. Such implicit descriptor-enabled ANN frameworks can guide the accelerated and tractable design of complex hierarchical materials for applications ranging from shock-absorbing layers in extreme environments to functional components in soft robotics.

\textbf{Keywords:} Architected materials, Artificial neural network, Structure-property relationship, Geometric descriptors, Multi-component shape invariants

\onehalfspacing

\doublespacing
\section{Introduction}

Hierarchically structured materials offer pathways to achieve optimal combinations of high energy absorption, high modulus, and high strength with simultaneous lightweighting. They overcome the limitations of conventional materials by leveraging superior intrinsic characteristics of structural features and their self and directed organizations across multiple length scales \cite{ritchie2011conflicts,wegst2015bioinspired, gupta2005nanoscale, fratzl2007nature,guo2023strong, buhrig2016biomimetic}. Hierarchically structured nanofibrous materials such as vertically aligned carbon nanotube (VACNT) foams have demonstrated an impressive combination of high energy absorption, modulus, and strength at low densities \cite{hutchens2011analysis,thevamaran2015shock}. Unlike viscoelastic polymeric foams, the VACNT foams dissipate energy through mechanisms that are independent of strain rate and temperature, making them ideal for a broad range of applications, especially those involving varying strain rates and temperatures \cite{xu2010carbon,yang2011modeling, gupta2024embracing}. These remarkable properties arise from their multi-lengthscale structural hierarchy, extending from nanoscale to macroscale, which can be controlled to achieve desired densities and mechanical properties by tailoring the synthesis process parameters during the VACNT synthesis \cite{lee2001temperature,chakrabarti2007number,zhang2008influence,raney2011tailoring}. 

\begin{figure}[htp]
	\centering
	\includegraphics[width=\linewidth]{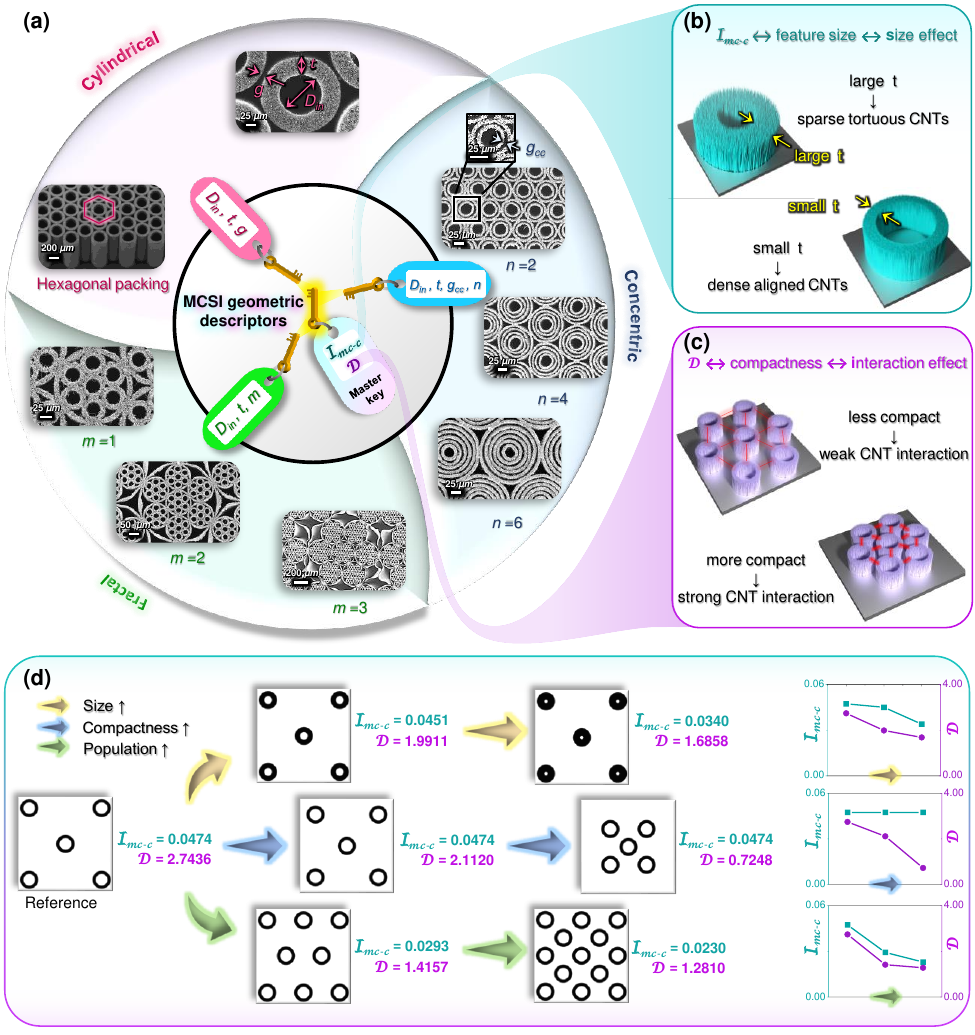}
	\caption{Implicit geometric descriptor based unified approach to design architectures. 
    (a)  Explicit geometric design parameters for hexagonally packed cylinder- (pink panel), concentric cylinder- (blue panel), and fractal- (green panel) arrays; MCSI implicit geometric descriptors---$\mathcal{I}_{mc-c}$ and $\mathcal{D}$---unifies the design of any arbitrary architectures (gray center panel) (b) $\mathcal{I}_{mc-c}$ captures the feature size that gives rise to size effect---more aligned and dense growth of CNTs within confined geometries.  
    (c) $\mathcal{D}$ captures the compactness that gives rise to interaction effect---enhanced inter-nanotube interaction in compact arrays. 
    (d) Illustration of the sensitivity of $\mathcal{I}_{mc-c}$ and $\mathcal{D}$ to the increment in feature size, population, and compactness within an architecture: $\mathcal{I}_{mc-c}$ predominantly captures the variation in size of the features whereas $\mathcal{D}$ predominantly captures the variation in compactness due to the variation in the feature size (yellow arrow);  $\mathcal{I}_{mc-c}$ is not sensitive to the variation in relative feature positions whereas $\mathcal{D}$ captures the variation in compactness due to the variation in relative positions of the features (blue arrow); $\mathcal{I}_{mc-c}$ captures the variation in feature population within the architecture whereas $\mathcal{D}$ captures the variation in the compactness due to the variation in the feature population (green arrow).}      
	\label{fig1}
\end{figure}

Synthesizing VACNTs in mesoscale architectures on a photolithographically prepatterned (chromium-coated) substrate is another effective approach to tailor the bulk density and mechanical properties of VACNT foams and achieve property amplifications in a non-additive fashion \cite{de2010diverse, de2011fabrication, lattanzi2014geometry,lattanzi2015dynamic, chawla2022superior, chawla2023disrupting, gupta2024embracing}. Recently, we demonstrated a synergistic enhancement of specific (density-normalized) mechanical properties---compressive modulus, strength, and energy absorption—in VACNT foams architected with hexagonally packed cylinders (Fig. \ref{fig1}(a)(pink)) compared to non-architected VACNT foams and other conventional foams \cite{chawla2022superior}. The enhancement in properties comes from a synthesis size effect---induced by the dense, aligned growth of CNTs in confined spaces (Fig. \ref{fig1}(b))---as well as the nanoscale interactions between the CNTs within and on the surface of the architectural elements (Fig. \ref{fig1}(c)). We further explored the design space by arranging cylindrical arrays into concentric (Fig. \ref{fig1}(a)(blue)) and fractal (Fig. \ref{fig1}(a)(green)) architectures while maintaining hexagonal packing of the unit cells. These configurations allow access to more favorable, linear density-dependent scaling of mechanical properties in architected VACNT foams to achieve better properties at ultra-low densities  \cite{chawla2023disrupting,gupta2024embracing}. These findings highlight the tremendous potential of architected VACNT foams to achieve tailored exceptional mechanical properties by harnessing intricate nanoscale mechanisms by design. 

Despite these progresses, the individual influences of the underlying nanoscale mechanisms---size effect and interaction effect---on the bulk mechanical properties of VACNTs remain elusive, limiting the ability to independently control these mechanisms to achieve target property combinations. Moreover, the current design approaches that rely on architecture-specific explicit geometric design parameters---inner diameter ($D_{in}$),  wall thickness ($t$), and gap between cylinders ($g$) in cylindrical arrays (Fig. \ref{fig1}(a)(pink)), gap between concentric rings ($g_{cc}$) and number of rings ($n$) in concentric cylinder arrays (Fig. \ref{fig1}(a)(blue)), and fractal order ($m$) in fractal arrays (Fig. \ref{fig1}(a)(green))---hinder the development of a unified structure-property relationship that broadly applies across different architectures.

To address this challenge, we introduce an artificial neural network (ANN) framework that correlates two multi-component shape invariant (MCSI) implicit geometric descriptors (see \cref{section:3.1}) (Fig. \ref{fig1}(a)(gray center) and Fig. \ref{fig1}(d)) \cite{rhouma2017moment,vzunic2018disconnectedness} of the architecture to the experimentally measured specific mechanical properties, such as specific energy absorption ($W^{*}$), specific peak stress ($\sigma^{*}$), and specific average modulus ($E^{*}$) to establish a unified structure-property relationship. These implicit geometric descriptors capture the size of the features within the architecture and the compactness of the architecture, which give rise to the nanoscale mechanisms that govern the bulk mechanical properties (Fig. \ref{fig1} (b-c)). Our implicit geometric descriptor-based approach enables both the prediction of mechanical properties of new architectures and the design of architectures for diverse mechanical property combinations through independent control of these descriptors. The findings of this study accentuates that the nanoscale mechanisms---such as size effects and inter-nanotube interactions---predominantly govern the mechanical properties of architected VACNT foams, rather than the specific spatial architecture. Unlike the complex deep learning models with numerous parameters, the reduced parameter space enabled by the MCSI implicit descriptors allows faster model training with much fewer experimental measurements---enabling efficient and accelerated design of architected materials where such multi-scale mechanisms interactively give rise to bulk mechanical behavior.

\section{Experimental methods}

\subsection{Synthesis of 2D architected VACNT foams} 

We synthesized the architected VACNT foams using floating catalyst thermal chemical vapor deposition (tCVD) \cite{fan2006single} process on patterned p-type silicon wafer with (100) crystal orientation. First, we spin coated the substrate with S1813 photoresist at 3000 rpm for 30 seconds  to a thickness of 10 $\mu m$. Then, we partially diced the 100 $mm$ diameter wafer to 30\% of its thickness (500 $\mu m$), creating a grid of 5 $mm$ $\times$ 5 $mm$ squares (A CAD file created in AutoCAD defined a grid of 5 $mm$ $\times$ 5 $mm$ squares on 4-inch substrate with each square containing the desired micropattern). We transfered the desired micropattern on to the photoresist-coated substrate using a Heidelberg DWL 66+ laser writer, which utilizes a 405 $nm$ UV laser diode for the process. Initially, the CAD file was converted into DXF format using ``x-convert" software and imported into the DWL 66 interface. We carefully mounted the coated substrate on the system’s holder while ensuring accurate alignment and leveling. A 10 $mm$ write head, capable of achieving a minimum feature size of ~1 $\mu m$, was used for the exposure. We calibrated the parameters such as laser power (60 $mW$), intensity (100\%), and filter (100\%) based on the properties of S1813 photoresist and silicon substrate. After exposure, we removed the unexposed photoresist by developing the wafer in an MF321 developer bath for 30 seconds. Next, we evaporated a chromium layer of 20 $nm$ at a rate of 0.5 \r{A}/$s$ onto the wafer using a metal evaporator. We then removed the UV-exposed photoresist in an acetone bath, leaving a chromium film with the inverse pattern of the desired VACNT architecture on the substrate that masked predefined areas preventing CNT growth. 

For tCVD, we injected a feedstock solution of ferrocene (catalyst precursor) dissolved in toluene (carbon source) into a carrier gas mixture of argon (760 $sccm$) and hydrogen (40 $sccm$) that was flown into a furnace tube at 0.8 $ml/min$ using a syringe pump. We maintained the furnace at 825°C (±5°C) and atmospheric pressure where VACNTs grow on the exposed silicon regions of the wafer. After synthesis, we removed the architected VACNT film from the furnace and separated into 5 $mm$ $\times$ 5 $mm$ squares. Each square had a specific combination of VACNT diameter, spacing, and lattice pattern, or sinusoid pattern defined by a particular sine function ready for subsequent mechanical characterization. We performed the required scanning electron microscopy (SEM) for the synthesized designs using a Zeiss Gemini 300 scanning electron microscope equipped with a field emission gun

All samples had a constant areal dimension of $5$ $mm$ $\times$ $5$ $mm$ with heights ($h$) ranging from 1 to 3 $mm$ (Tables S2-S4).  We designed 81 different architectures that includes 60 cylindrical architectures \cite{chawla2022superior}, 18 concentric cylindrical architectures \cite{chawla2023disrupting}, and 3 fractal architectures \cite{gupta2024embracing}. We designed the 60 cylindrical architectures using a full factorial design of experiments (DOE) approach, varying the explicit geometric design parameters, $D_{in}$, $t$ and $g$. To create concentric cylindrical architectures, explicit design parameters, $D_{in}$, $t$, $g_{cc}$ and $n$ were varied, keeping the $D_{in}/t$ ratio at 5, while for fractal arrays, $m$ was varied from 1 to 3. Table \ref{T1}  summarizes the different levels of explicit design parameters used to design the cylindrical, concentric cylindrical, and the fractal arrays. 

To test and validate our ANN models on designs that go beyond these regular cylinder-based architectures described by their respective explicit geometric design parameters, we designed additional cylinder-based architectures (Fig. \ref{fig5}(a-d)) and sinusoid-based architectures (Fig. \ref{fig5}(e-i))) that have different explicit geometric design parameters. We measured the mass of individual samples using analytical micro-balance and calculated the volume from dimensions measured using precision calipers ($5 mm \times 5 mm \times h$) (Tables S2-S4 and Table S5). We then determined the bulk density of the architected samples ($\rho$) by dividing their mass by the volume.
\begin{table}[htp]
\centering
 \caption{Explicit design parameters and their different levels used to create different hexagonally packed cylindrical, concentric and fractal architectures}
 \vspace{0.5 cm}
  \begin{tabular}[htbp]{@{}lll@{}}
    \hline
    Architecture & Explicit design parameter & Levels ($\mu m$) \\
    \hline
        Cylindrical  &$D_{in}$  & 50, 100, 200  \\
    {}  & $g$  & 0, 20 50, 100, 200  \\
    {}  & $t$ & 10, 20, 40, 100  \\
    Concentric  & $D_{in}\textsuperscript{a}$  & 25, 50  \\
    {}  & $t$  & 5, 10  \\
    {}  & $g_{cc}$  & 5, 20, 50  \\
    {}  & $n$  & 2, 4, 6  \\
    Fractal  & $D_{in}\textsuperscript{b}$  & 25
    \\
    {}  & $t$  & 5  \\
    {}  & $m$  & 1, 2, 3  \\
       
    \hline
  \end{tabular}
  \label{T1}\\
  \raggedright
  \textsuperscript{a)} Here $D_{in}$ implies the inner diameter of the innermost cylinder in the concentric array.\\
  \raggedright
  \textsuperscript{b)} Here $D_{in}$ implies the inner diameter of cylinders of $1^{st}$ fractal order.
\end{table}

\begin{figure}[htp]
	\centering
	\includegraphics[width=\linewidth]{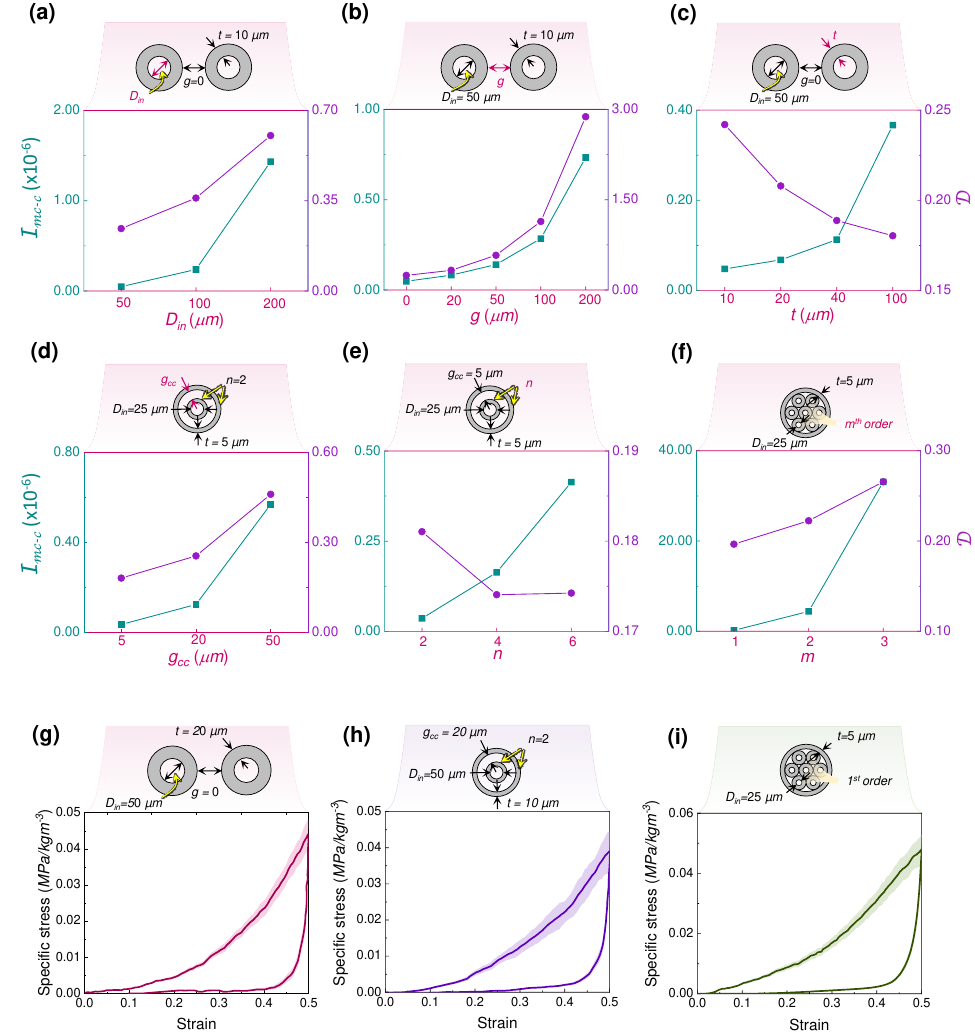}  
	\caption{Variation of implicit geometric descriptors---$\mathcal{I}_{mc-c}$ (cyan) and $\mathcal{D}$ (violet)---across different architectures as a function of their explicit design parameters (a)  $D_{in}$ in cylindrical architecture, (b) $g$ in cylindrical architecture, (c) $t$ in cylindrical architecture, (d) $g_{cc}$ in concentric cylindrical architecture, (e) $n$ in concentric cylindrical architecture and (f) $m$ in fractal arrays. (g-i) Representative specific stress-strain profiles for each architecture type, with shaded regions indicating standard deviation in the measured specific stress: (g) for cylindrical \cite{chawla2022superior}, (h) for concentric cylindrical \cite{chawla2023disrupting} (i) for fractal array \cite{gupta2024embracing} (Top panels illustrate the explicit design parameters)}
	\label{fig2}
\end{figure}

\subsection{Mechanical testing of architected VACNT foams and calculation of their specific mechanical properties}   
We measured the mechanical properties of all the architected VACNT foams used in this work using uniaxial compression experiments at a quasistatic strain rate of 0.01 $s^{-1}$\cite{chawla2022superior,chawla2023disrupting,gupta2024embracing}. We performed these experiments using an Instron E3000 Electropulse universal testing system---designed to accommodate soft materials---employing platens and fixtures carefully matched to both machine and sample compliance, as well as to the specific sample geometry. The system was equipped with a 5 $kN$ load cell and was controlled through the Instron software, ``wavematrix". We controlled the maximum compressive strain and the strain rate by prescribing the displacement and displacement rate using the wavematrix. The specific mechanical properties of interest---$W^{*}$, $\sigma^{*}$ and $E^{*}$---were calculated from the quasistatic compressive stress-strain curves (Fig. \ref{fig2}(g-i) and the bulk densities of the architected samples using Equations \ref{eq1}, \ref{eq2}, and \ref{eq3} respectively:

\begin{equation}
    W^{*} = \frac{1}{\rho}\int_{A} \sigma (\epsilon) d\epsilon
    \label{eq1}
\end{equation}

\begin{equation}
   \sigma^{*}= \frac{\sigma|_{\epsilon=\epsilon_{max}}}{\rho}
    \label{eq2}
\end{equation}    

\begin{equation}
      E^{*}= \frac{2}{\rho\epsilon_{max}^2}\int_{\epsilon=0}^{\epsilon=\epsilon_{max}} \tilde{\sigma} (\epsilon) d\epsilon
    \label{eq3}
\end{equation}

Here, $\epsilon$ is the strain, obtained by normalizing the displacement by the sample height ($h$). $\sigma$ is the stress, determined by dividing the measured force by the sample's area.  $\epsilon_{max}$ is the maximum strain, which is 0.5 in this study. 

Equation \ref{eq1} for $W^*$ represents the area under the loading-unloading hysteresis loop, while Equation \ref{eq2} for $\sigma^*$ represents the stress measured at $\epsilon_{max}$. The $E^{*}$ in Equation \ref{eq3} is calculated by integrating the mean stress $\tilde{\sigma}$---averaged over the loading and unloading branches---up to $\epsilon_{max}$ and then normalizing by $\epsilon_{max}$. We adopt this integral modulus, rather than the conventional tangent modulus calculated from the initial loading curve, for two key reasons. First, architected VACNT foams, especially those designed for protective applications, typically operate under large deformations where the stress–strain response is nonlinear and continuously evolving. More importantly certain coarse architectures---such as concentric cylinders with large $g_{cc}$---undergo transition from shell buckling to a structural buckling mode, introducing sub‑linear trends (Fig. \ref{fig5}(j-l)) in the stress-strain response at intermediate strains \cite{chawla2023disrupting}. This response further evolves at higher strains where the nanoscale mechanisms such as size effect and nanoscale interactions become more dominant. Therefore, an average modulus estimated over a large deformation range  offers a more representative measure of the foam's overall response relevant to protective applications. Second, VACNT foams primarily dissipate energy through friction arising from inter-nanotube interactions. This affects not only the loading behavior but also the unloading behavior---that also contributes to the material’s dissipation of the stored elastic energy. Therefore, by calculating the average modulus using the mean stress over both loading and unloading segments, we capture a more accurate representation of the foam’s elastic energy storage \cite{chawla2023disrupting}. 
 
\section{Results and discussion}
\subsection{MCSI implicit geometric descriptors unify the structure-property relationship across different architectures}
\label{section:3.1}
 
\begin{figure}[t]
	\centering
	\includegraphics[width=\linewidth]{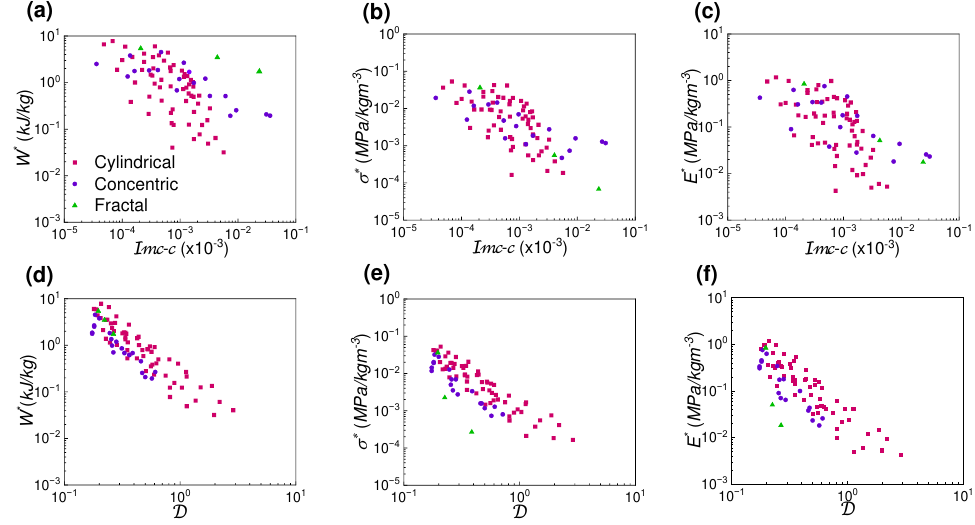}  
	\caption{Variation of experimentally measured specific mechanical properties of different architectures as functions of implicit geometric descriptors.(a) $W^*$ vs. $\mathcal{I}_{mc-c}$. (b) $\sigma^*$ vs. $\mathcal{I}_{mc-c}$. (c) $E^*$ vs. $\mathcal{I}_{mc-c}$. (d) $W^*$ vs. $\mathcal{D}$. (e) $\sigma^*$ vs. $\mathcal{D}$. (f) $E^*$ vs. $\mathcal{D}$}
	\label{fig3}   
\end{figure}

Explicit geometric parameters are inherently architecture‑dependent: for example, $D_{in}$ is 
 well defined for cylinder-based designs where it explicitly describes the inner diameter of cylinders, however it's meaning becomes ambiguous in non‑cylindrical designs such as sinusoidal and other arbitrary designs. In addition, each such explicit geometric parameter describes only an isolated geometric feature of the architecture, thereby omitting the collective geometric attributes that primarily govern the general bulk mechanical behavior across various architectures. These limitations hinder the development of a truly universal structure–property relationship for architected VACNT foams. By contrast, implicit geometric descriptors that are both architecture‑agnostic and describe global features of the architecture that give rise to the key phenomena---universally governing the bulk mechanical behavior across various architectures---provide a more robust basis to model the unified structure-property relationship. 

In our previous work, we showed that the mechanical properties of architected VACNT foams are primarily influenced by two key nanoscale phenomena. The first is the tendency of VACNTs to grow more densely and align better within geometrically confined regions of the architecture 
 (Fig. \ref{fig1}(b)) \cite{chawla2022superior,chawla2024superior}. This creates a thickness-dependent size effect as we have seen in cylindrical architectures, where cylinders with smaller wall thicknesses result in higher specific mechanical properties compared to cylinders with larger wall thicknesses \cite{chawla2022superior}. The second influential mechanism is the interaction among nanotubes at the nanoscale. This phenomena, particularly in architectures with compact geometries without significant gap between adjacent cylinders, increases the strength and modulus of the material due to intensified nanotube interactions and enable greater energy dissipation through friction between neighboring nanotubes during deformation (Fig. \ref{fig1}(c)) \cite{chawla2022superior}. Building on these insights, we aim to capture both the size effect and the interaction effect on the mechanical properties by employing implicit geometric descriptors that are sensitive to the size and compactness of the geometric features in the architecture. By correlating these descriptors to the mechanical properties of various types of 2D architectures, we establish a unified structure-property relationship that applies to broad classes of 2D architected VACNT foams with different sets of explicit geometric design parameters.

We selected our geometric descriptors for 2D architectures from image classification studies using MCSIs---invariant to translation, rotation and scaling of the multi-component shape---introduced by Rhouma et al. \cite{rhouma2017moment} and Zunic et al. \cite{vzunic2018disconnectedness}. These MCSIs are based on Hu's geometric moment invariants \cite{hu1962visual} (Equation \ref{eq7}) and facilitate shape-based object analysis. For our application, we computed the MCSIs by treating the entire 5 $mm$ $\times$ 5 $mm$ architectural design as a multi-component shape to ensure consistent and unbiased evaluations (see Supplementary Information). 

Among the descriptors introduced by Rhouma et al. \cite{rhouma2017moment}—$\mathcal{I}_{mc-a}$, $\mathcal{II}_{mc-a}$, $\mathcal{I}_{mc-c}$, and $\mathcal{II}_{mc-c}$—the first two are sensitive exclusively to feature size, while $\mathcal{I}_{mc-c}$ and $\mathcal{II}_{mc-c}$ are sensitive to both the size and population of features within the architecture. We prioritized $\mathcal{I}_{mc-c}$ and $\mathcal{II}_{mc-c}$ due to their ability to capture both the effect of architectural feature size and their population within the  5 $mm$ $\times$ 5 $mm$  area. It is noteworthy that $\mathcal{I}_{mc-c}$ is derived from the first-order Hu's moment invariant, offering more generalization capability, whereas $\mathcal{II}_{mc-c}$, derived from the second-order invariant, is more suitable for capturing finer geometric intricacies. Given our focus on generalization and simplicity in establishing the structure-property relation, we selected $\mathcal{I}_{mc-c}$ (Equation \ref{eq9}) as our first implicit geometric descriptor.

Similarly, in another study, Zunic et al. \cite{vzunic2018disconnectedness} derived the MCSI, $\mathcal{D}$, that measures the disconnectedness---essentially the inverse of compactness---of the features within a multi-component shape. We utilized $\mathcal{D}$ (Equation \ref{eq10}) as our second descriptor to describe the effective compactness of our 2D architecture. Both $\mathcal{I}_{mc-c}$ and $\mathcal{D}$ are continuous, real-valued variables defined on the interval (0,$\infty$). The definitions and the corresponding derivations for the geometric descriptors  $\mathcal{I}_{mc-c}$ and $\mathcal{D}$ are given below. Comprehensive derivations can be found in \cite{rhouma2017moment} and \cite{vzunic2018disconnectedness}.

The moment $m_{p,q}(S)$ of a given planar shape $S$ is given as,

\begin{equation}
    m_{p,q}(S)= \int_{y}\int_{x}x^py^qdxdy
\end{equation}

The order of the moment is given by \( p + q \). The $0^{th}$ order moment, where \( p = q = 0 \), represents the area of \( S \), while the first-order moments can be used to determine the centroid of \( S \). The centroid of $S$ is given by,

\begin{equation}
    (x_c(S),y_c(S)) = \left( \frac{m_{1,0}(S)}{m_{0,0}(S)}, \frac{m_{0,1}(S)}{m_{0,0}(S)}\right)
\end{equation}

Centralized moments are calculated by taking the moments relative to the centroid, as given by the following expression,

\begin{equation}
   \overline{m}_{p,q}(S)= \int_{y}\int_{x}(x-x_c(S))^p(y-y_c(S))^qdxdy
\end{equation}

For a given shape $S$, the first Hu's geometric moment invariant is given by,

\begin{equation}
\mathcal{H}_1(S) = \frac{1}{m_{0,0}(S)^2}.(\overline{m}_{2,0}(S)+\overline{m}_{0,2}(S))
\label{eq7}
\end{equation}

If $S$ is a multi-component shape  composed of $n$ components $(S_1,S_2,...S_n)$, then 

\begin{equation}
S = S_1\cup S_2 \cup....\cup S_n
\end{equation}

Based on the above definitions, the MCSIs---$\mathcal{I}_{mc-c}(S)$ and $ \mathcal{D}(S)$---of the multi-component shape, $S$ can be derived in the following forms, 

\begin{equation}
    \mathcal{I}_{mc-c}(S) = \frac{1}{{m_{0,0}(S)}^2}\sum_{i=1}^{n}(\overline{m}_{2,0}(S_i)+\overline{m}_{0,2}(S_i))
    \label{eq9}
\end{equation}

\begin{equation}
    \mathcal{D}(S) = \mathcal{H}_1(S)-\frac{1}{{m_{0,0}(S)}^3}\sum_{i=1}^{n}m_{0,0}(S_i)^3.\mathcal{H}_{1}(S_i)
    \label{eq10}
\end{equation}

Fig. \ref{fig1}(d) illustrates how the descriptors $\mathcal{I}_{mc-c}$ and $\mathcal{D}$ effectively capture variations in size, population, and compactness of features within a 2D architecture. As it can be seen, $\mathcal{I}_{mc-c}$ is sensitive solely to the relative size and population of the features within the architecture, unaffected by their spatial positions. $\mathcal{D}$ captures compactness of the architecture resulting from all changes in position, size, and population of the components within the architecture. 

Fig. \ref{fig2} shows the variation of implicit geometric descriptors---$\mathcal{I}_{mc-c}$ (cyan) and $\mathcal{D}$ (violet)---with various explicit geometric design parameters---$D_{in}, t, g, g_{cc}, n$, and $m$--- across different architectures. These clear trends collectively highlight that the chosen geometric descriptors, $\mathcal{I}_{mc-c}$ and $\mathcal{D}$, are fundamentally implicit functions of these conventionally used explicit geometric design parameters. This demonstrates the ability of selected geometric descriptors to encapsulate the geometric intricacies of a wide range of architectures into a single variable that facilitates unifying the structure-property relation across different architectural designs. To assess the potential of our selected geometric descriptors in capturing the specific mechanical properties of the architected VACNT foams, we plot each of the specific mechanical properties---$W^*$, $\sigma^*$, and $E^*$---as a function of each of the two geometric descriptors, $\mathcal{I}_{mc-c}$ and $\mathcal{D}$ in Fig. \ref{fig3}. These plots demonstrate that both descriptors, $\mathcal{I}_{mc-c}$ and $\mathcal{D}$, individually exhibit strong nonlinear correlation with the specific mechanical properties, and demonstrate their suitability as predictors for our ANN model.
\begin{figure}[htp]
	\centering
	\includegraphics[width=\linewidth]{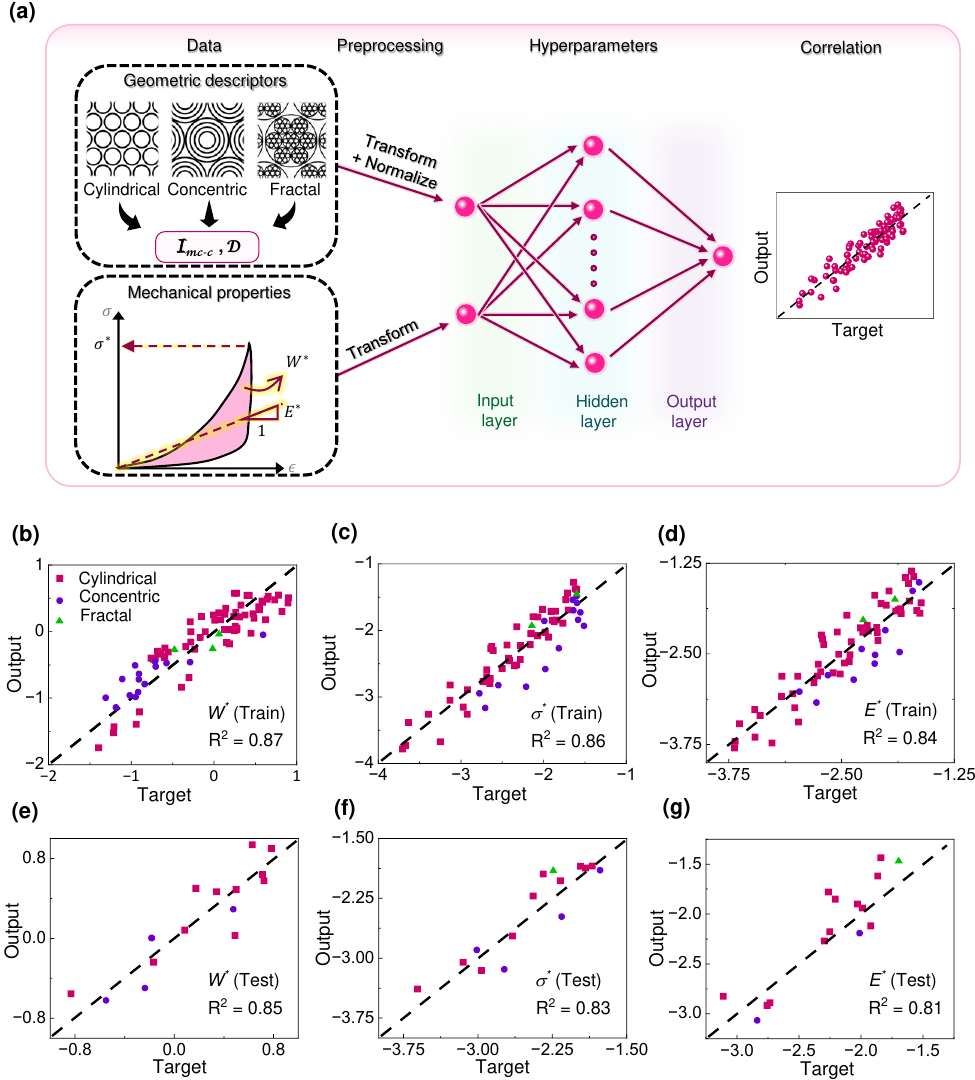}  
\caption{Implicit geometric descriptor-enabled ANN framework for the unified structure property relationship. (a) Implicit geometric descriptor-enabled ANN framework. Coefficient of determination for (b) training split for $W^*$, (c) training split for $\sigma^*$, (d) training split for $E^*$, (e) testing split for $W^*$, (f) testing split for $\sigma^*$ and (g) testing split for $E^*$ }
\label{fig4}
\end{figure}

\subsection{Development of the ANN framework to establish unified structure-property
correlations}       
ANNs are used effectively in various fields, including materials science and mechanics, for uncovering complex, non-trivial structure-property relationships that traditional models often struggle to capture \cite{guo2021artificial,masi2021thermodynamics,ma2020accelerated,liu2020machine,moghadam2019structure}. They are also well known for their flexibility in processing diverse types of data.  Conversely, convolutional neural networks (CNNs) are mainly designed for grid-like data, such as images, and are commonly used to identify geometric patterns in 2D or 3D datasets to find structure-property relationships based on topological and morphological features. These CNNs typically require large datasets---usually generated through simulations or synthetic data creation techniques---to identify robust patterns \cite{abueidda2019prediction, yang2020prediction}. However, such large datasets can lead to significant computational demands and increased training times. Here, we employ an efficient ANN approach that utilizes only 81 architectures, where the two key geometric descriptors---$\mathcal{I}_{mc-c}$ and $\mathcal{D}$---computed for each architecture served as predictors, while the experimentally measured mechanical properties for each architecture served as target observations (Fig. \ref{fig4}(a)).      

Choosing the right ANN model is essential as it significantly affects the model's performance, computational efficiency, and ability to generalize to unseen data outside training set \cite{abiodun2018state}. Multilayer perceptron (MLP) is one of the most recognized and commonly used ANN models  \cite{taud2018multilayer}. This is a feed-forward ANN that utilizes the backpropagation technique for learning. The structure consists of an input layer of neurons that receives the input data, one or more hidden layers that iteratively process the data, and an output layer that generates the predicted output. Here, the cost function is calculated and minimized by comparing the network's output with the true target value. We employed a MLP using MATLAB for our framework.

Leven-Marquardt (LM) \cite{more2006levenberg} and Bayesian regularization (BR) \cite{burden2009bayesian} are the commonly used optimization techniques to train ANNs with relatively small datasets. While LM is often used for faster convergence in nonlinear least-squares problems, it lacks emphasis on regularization, which is important for avoiding overfitting, especially with limited data (Figure S2). To enhance the model’s generalization and robustness to unseen data, we opted for the BR backpropagation algorithm, which reduces model complexity by regularizing the cost function \cite{kayri2016predictive,khan2020design,payal2013comparative}. Unlike other early stopping-based methods, this method does not require a validation set to monitor performance \cite{payal2013comparative}, instead it incorporates prior knowledge about weight distribution and penalizes complex models for unreasonable assignment of weights, leading to better generalization (see Supplementary Information and Figure S2). Therefore, BR is also a generally preferred regularization technique--- particularly with smaller datasets---over traditional regularization techniques such as L1 and L2 regularization methods, that rely on cross‑validated hyperparameters and dropout methods that often underfit with small datasets \cite{zhang2021understanding,fila2024mitigating}.

We used the mean square error (MSE) (Equation \ref{eq11}) as the cost function: 

\begin{equation}
\mathrm{MSE} = \frac{1}{N} \sum_{i=1}^{N}(Y_{i} - \hat{Y}_{i})^2
\label{eq11}
\end{equation}
Here, $Y_{i}$ and $\hat{Y}_{i}$ denote the predicted and target output variables, where $N$ is the number of observations. There was no need to manually set a threshold for the cost function, as BR automatically adjusts parameters like regularization terms and stopping criteria based on the data and model complexity.
 
Selecting the appropriate data preprocessing technique can greatly improve ANN performance, particularly when dealing with a limited dataset \cite{millan2021application,yu2021deep,srinivasan2024generalized}. Methods such as logarithmic transformation \cite{millan2021application}, which helps to linearize relationships, and normalization \cite{yu2021deep,srinivasan2024generalized}, which standardizes the scale of input features, enable the network to better capture essential patterns and minimize noise. 

Before training, we applied a $log_{10}$ transformation to preprocess the input data (Equations \ref{eq12}-\ref{eq13}):
\begin{equation}
    X^\prime = log_{10}(X)
    \label{eq12}
    \end{equation}
    \begin{equation}
    Y^\prime = log_{10}(Y)
     \label{eq13}
      \end{equation}
 Here, $X$ and $Y$ represent the predictor and target data respectively, while $X'$ and $Y'$ correspond to their transformed versions. Additionally, Z-score normalization (Equation \ref{eq14}) was applied to the input descriptors:
         \begin{equation}
    Z = \frac{X^\prime-\mu}{\sigma}
    \label{eq14}
\end{equation}

\begin{figure}[htp]
	\centering
	\includegraphics[width=0.91\linewidth]{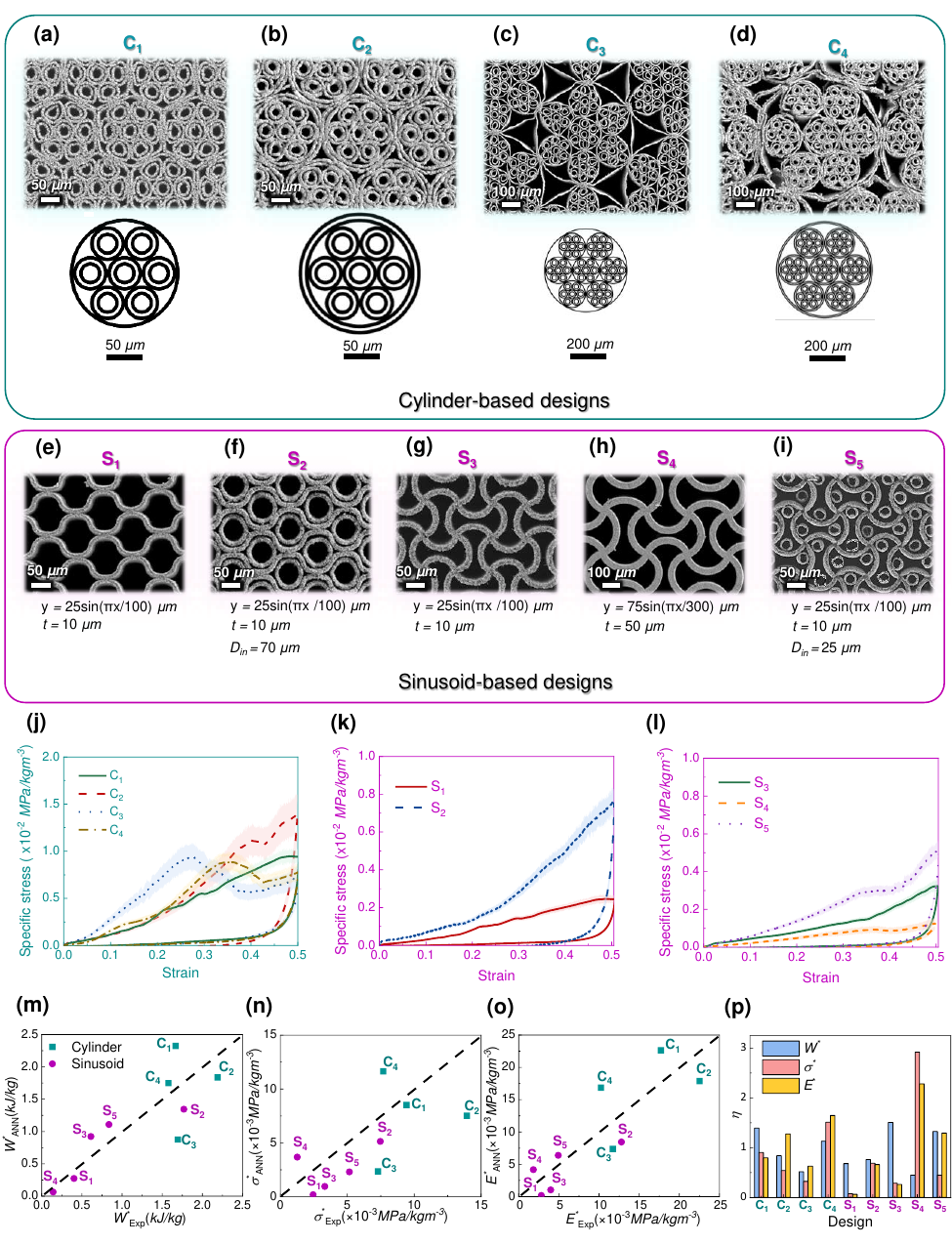}
	\caption{Generalizability of ANN models to new cylinder-based and sinusoid-based designs: (a-d) SEM images of cylinder-based fractal designs created by incorporating concentric rings, with corresponding unit cells shown below: (a) $C_1$, (b) $C_2$, (c) $C_3$, and (d) $C_4$.
    (e-i) SEM images of sinusoid-based designs with corresponding sine function, thickness and diameter of the inscribed cylinder given below: (e) $S_1$, (f) $S_2$, (g) $S_3$, (h) $S_4$, and (i) $S_5$. Representative specific stress-strain profiles, with the shaded regions indicating the standard deviation in the measured specific stress for (j) cylinder-based designs, (k) one-way sinusoid based designs and (l)  two-way sinusoid-based designs. Comparison of values predicted by ANN with the experimentally measured values across architectures, $C_1$ to $S_5$ for (m) $W^*$, (n) $\sigma^*$ and (o) $E^*$. (p) Ratio ($\eta$) of ANN-predicted to the experimentally-measured values for $W^*$, $\sigma^*$ and $E^*$ across $C_1$ to $S_5$ }

	\label{fig5}
\end{figure}

Here, $\mu$ and $\sigma$ denote the mean and the standard deviation of the transformed descriptor space. This 
preprocessing technique proved highly effective in reducing the learning burden on the network, allowing it to model the structure-property relationship more efficiently. The performance of the neural network is also influenced by the number of hidden layers and the number of neurons in each hidden layer, making it essential to adjust these two hyper parameters appropriately \cite{taud2018multilayer}. With the employed data preprocessing technique, the optimal configuration for the ANN required one hidden layer consisting of 13, 19, and 21 neurons for $ W^*$, $\sigma^*$, and $E^*$ respectively. We found sigmoid function (Equation \ref{eq15}) as the best performing nonlinear activation function to capture complex patterns in our data for all three mechanical properties: 
\begin{equation}
f(x) = \frac{1}{1 + e^{-x}}
\label{eq15}
\end{equation}
For each mechanical property, the ANN differed only in the optimal number of neurons in the hidden layer. The grid search-based parameter optimization process (Figure S1) that we employed is illustrated in Supplementary Information . We split the input dataset of 81 designs into two subsets, 66 designs for training, and 15 designs for testing. Using a holdout validation approach (shuffle-split procedure) \cite{may2010data,yu2021deep}, we trained the network multiple times with different shuffled splits, then averaged the performance metrics to obtain a more reliable estimate of the model’s overall performance.

The ANN model demonstrated excellent correlations for all three mechanical properties under consideration, indicating its effectiveness in capturing the structure-property relationships. As shown in Fig. \ref{fig4}(b-g), the coefficient of determination ($R^2$) greater than 0.8 and mean absolute percentage errors (MAPE) around 10-15 \% (Table \ref{T2}) for each property highlight the model's impressive performance after training. These results underscore the capability of the implicit geometric descriptor-based ANN approach to successfully establish a unified structure-property relationship that integrates various types of 2D architectures. We compared the performance of our ANN models with other regression models (see Supplementary Information), including multivariate linear regression (MVLR), MVLR with an added interdependence term given by the product of the descriptors (MVLR-I) (Equation S1) and Gaussian process regression (GPR) (Equation S2). Table \ref{T2} provides the summary of the $R^2$ and MAPE obtained with each model and demonstrates that the ANN models outperform other regression models in establishing accurate correlations. The relatively strong performance of other regression models also highlight the effectiveness of the implicit geometric descriptor-based approach in capturing the hidden structure-property correlations. Our ANN's ability to establish a unified structure-property relationship using implicit geometric descriptors that capture the size effect and interaction effect suggests that, in architected VACNT foams, mechanisms like size effect and interaction effect are the primary drivers of bulk mechanical properties, irrespective of the specific 2D architecture---a finding that underscores the utility of the geometric descriptors in ANN rather than mere image regression-based approaches.

\begin{table}[htp]
\centering
\caption{Comparison of $R^2$ and MAPE across various regression models}
\vspace{0.5 cm}
\begin{tabular}{lllllll}        
\hline
Model & \multicolumn{2}{c}{$W^*$} & \multicolumn{2}{c}{$\sigma^*$} & \multicolumn{2}{c}{$E^*$} \\ \cline{2-7} 
      & $R^2$    & MAPE        & $R^2$    & MAPE        & $R^2$    & MAPE   \\     
      & {}    &  (\%)       & {}    &  (\%)       & {}    &  (\%)
      \\ \hline
MVLR  & 0.69 & 29.55           & 0.59 & 37.54           & 0.58 & 39.88           \\ 
MVLR-I & 0.77 & 22.34          & 0.74 & 24.66           & 0.69 & 26.57           \\ 
GPR   & 0.81 & 19.22           & 0.77 & 21.12           & 0.76 & 20.64           \\ 
ANN   & 0.87 & 13.14           & 0.86 & 13.66           & 0.84 & 14.90           \\ \hline
\end{tabular}
\label{T2}
\end{table}

\begin{figure}[htp]
	\centering
	\includegraphics[width=\linewidth]{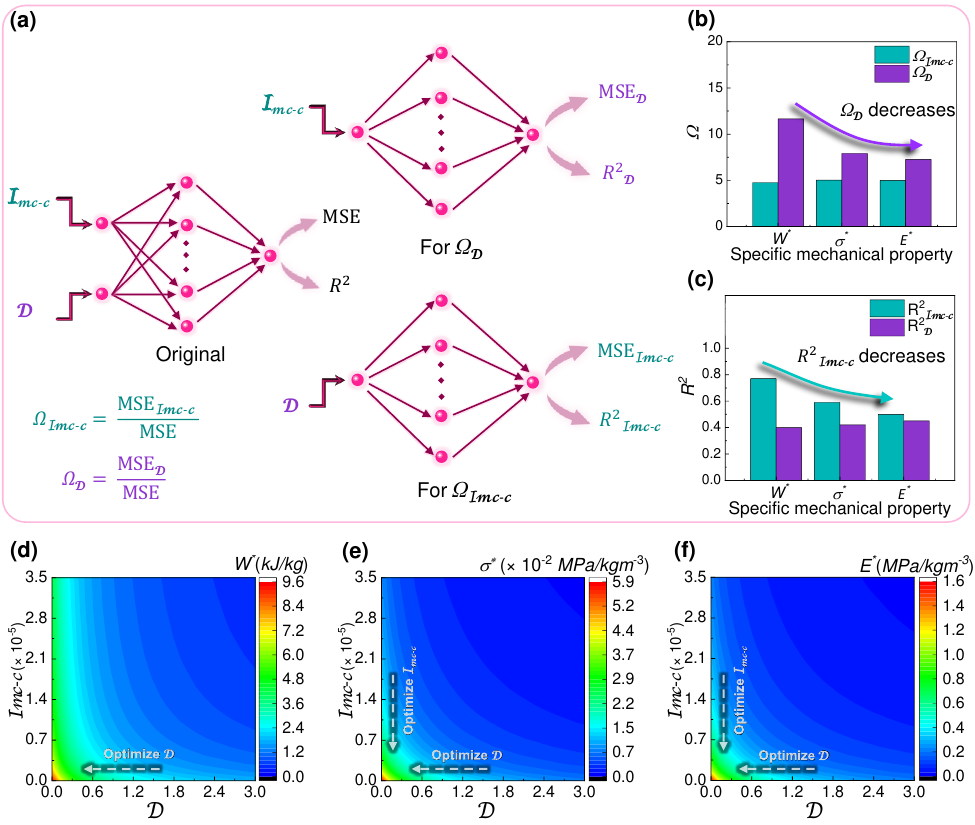}
	\caption{
    Design of architected materials with implicit geometric descriptors. (a) Schematic illustration of the various training conditions used to evaluate the descriptor importance scores---$\Omega_{\mathcal{I}_{mc-c}}$ and $\Omega_{\mathcal{D}}$---and the coefficients of determination---$R^2_{\mathcal{I}_{mc-c}}$ and $R^2_{\mathcal{D}}$---for the descriptors $\mathcal{I}_{mc-c}$ and $\mathcal{D}$, respectively. (b) Comparison of the descriptor importance scores---$\Omega_{\mathcal{I}_{mc-c}}$ and $\Omega_{\mathcal{D}}$---demonstrating the influence of each descriptor on specific mechanical properties $W^*$, $\sigma^*$, and $E^*$; lower values of $\Omega_{\mathcal{D}}$ in spite of $\Omega_{\mathcal{I}_{mc-c}}$ not changing significantly implies the interdependence between the descriptors in influencing $\sigma^*$ and $E^*$. (c) Comparison of the coefficients of determination---$R^2_{\mathcal{I}_{mc-c}}$ and $R^2_{\mathcal{D}}$---showing similar trend as in the case of descriptor importance scores. Design contours for the following specific mechanical properties as functions of implicit geometric descriptors predicted by the ANN model: (d) $W^*$, (e) $\sigma^*$ and (f) $E^*$}
	\label{fig6}             
\end{figure}

\subsection{Generalization capability of implicit geometric descriptor-enabled ANN framework}

To demonstrate that the unified structure property relationship we developed could serve as a predictive model to predict the mechanical properties of previously unknown architectures, we designed several new architectural designs---both cylinder-based and sinusoid-based---ensuring they were distinct from the 81 architectures used to train the ANN. However, we kept the descriptor values of these new architectures within the range of those 81 architectures used in the training of the ANN. The new cylinder-based designs were created by incorporating concentric rings within fractal arrays, resulting in four distinct architectural types ($C_1-C_4$)(Fig. \ref{fig5}(a-d)): $C_1$ and $C_2$ are first-order fractal arrays where $C_1$ has concentric rings applied to the inscribed cylinders while $C_2$ has concentric rings applied to both the inscribed and circumscribed cylinders. $C_3$ and $C_4$ are second-order fractal arrays where $C_3$ has concentric rings applied to the first-order inscribed cylinders while  $C_4$ has concentric rings applied to both the first- and second-order inscribed cylinders as well as the circumscribed cylinder. The sinusoid-based designs, included 5 different types ($S_1-S_5$) (Fig. \ref{fig5}(e-i)): $S_1$ and $S_2$ have sinusoids running in single direction while $S_3$, $S_4$, and $S_5$ employ orthogonal sinusoid patterns with sinusoids running in both directions. $S_2$ has the same sinusoidal pattern as $S_1$ with additional cylinders in the gaps, and $S_5$ has additional cylinders within the gaps of $S_3$. $S_4$ differs from $S_3$, using a different sine function for its design. Fig. \ref{fig5}(j–l) presents the representative specific stress–strain curves for these cylinder-based designs ($C_1$–$C_4$) and the sinusoid-based designs ($S_1$–$S_5$).

The ANN models gave accurate predictions for all three specific mechanical properties for the architectures---$C_1$ to $S_5$---used to demonstrate it's predictive capability. In Fig. \ref{fig5}(m-o), the predicted values of these mechanical properties are compared with experimental values, demonstrating the model's accuracy. Fig. \ref{fig5}(p) summarizes this by presenting the ratio ($\eta$) of predicted to experimental values for $W^*$, $\sigma^*$, and $E^*$ across designs $C_1$ to $S_5$. Notably, the prediction performance for the sinusoid-based architectures was not only excellent but slightly better than that of most of the cylinder-based ones. This is a result of the descriptor values of sinusoid-based architectures being closer to the expected values of the distribution of the descriptors used during training. This outcome suggests that our ANN-enabled structure property relationship---based on implicit geometric descriptors---is generalizable across a wide spectrum of architectures. 

The minor deviations in the predictions highlight the higher order effects emerging from the architectural intricacies---subtleties that extend beyond the first-order nanoscale mechanism driven trends captured by the two MCSI descriptors employed in this framework. To account for this, the flexibility of our framework allows for straightforward refinements: expanding the dataset to include a richer variety of architectural designs, as well as incorporating higher‑order geometric moment‑derived MCSIs that also encode local architectural details. These refinements would enable the model to bridge the gap between universally governing nanoscale mechanisms and higher order effects due to architecture-specific nuances, ultimately delivering high-precision predictions across a broad range of nanofibrous foam architectures.

\subsection{Implicit geometric descriptor-based design for independent tuning of mechanical properties}

To guide the implicit geometric descriptor-based design of architected VACNT foams, as a first step, we quantified the individual influence of each descriptor on the mechanical properties, through a ``quotient of error increase" sensitivity analysis \cite{mrzyglod2020sensitivity} for our ANN model. This method measures how much the network's error increases when a specific descriptor is excluded, compared to when all descriptors are used. It enables us to assess the importance of each descriptor by evaluating the impact of its absence on the model's overall performance. For this analysis, we trained the model separately for each mechanical property of interest and computed the MSE and $R^2$ values using only one of the two geometric descriptors (Fig. \ref{fig6}(a)). The descriptor importance score $\Omega_\mathcal{X}$ is the ratio between the MSE obtained when the descriptor $\mathcal{X}$ is excluded---$\mathrm{MSE}_{\mathcal{X}}$---and the MSE for the original network (Equation \ref{eq16}), providing a measure of the model's sensitivity to $\mathcal{X}$.
Comparing $\Omega$ values across descriptors reveals their relative importance to the ANN model, as shown in Fig. \ref{fig6}(b).
\begin{equation}
    \Omega_\mathcal{X} = \frac{\mathrm{MSE}_{\mathcal{X}}}{\mathrm{MSE}}
    \label{eq16}
\end{equation}

The consistently higher values obtained for $\Omega_\mathcal{D}$ across all three specific mechanical properties indicates that $\mathcal{D}$ is the more influential parameter of the two in determining the mechanical properties (Fig. \ref{fig6}(b)). This suggests that, in general, the compactness of the geometries in the architecture should be prioritized to achieve higher specific mechanical properties. However, notably, the influence of $\mathcal{D}$ is lower on $\sigma^*$ and even lesser on $E^*$, despite $\Omega_{\mathcal{I}_{mc-c}}$ remaining almost unchanged (Fig. \ref{fig6}(b)). This trend is also reflected in the $R^2$ values computed for the ANN when one of the geometric descriptors is excluded: the $R^2$ values are consistently higher in the absence of $\mathcal{I}_{mc-c}$, although they are reduced for $\sigma^*$ and $E^*$ (Fig. \ref{fig6}(c)). This finding indicates that $W^{*}$ is predominantly governed by the nanoscale interactions among the nanotubes---captured by $\mathcal{D}$---whereas there is a strong interplay between size effect---captured by $\mathcal{I}_{mc-c}$---and the nanotube interactions in governing the mechanical properties $\sigma^*$ and $E^*$. Our multi-variate regression analysis with interdependence term included (MVLR-I) further supports this statistically significant interdependence between $\mathcal{D}$ and $\mathcal{I}_{mc-c}$, with p-values of 0.00027 for $\sigma^*$ and 0.00015 for $E^*$.  These findings imply that densely packed and aligned CNTs within confined regions enhance nanotube interactions upon loading, reinforcing the critical role of compactness on $\sigma^*$ and $E^*$. Therefore, optimizing both $\mathcal{I}_{mc-c}$ and $\mathcal{D}$ in the design of architected VACNT foams could enhance strength and modulus, while designs focused solely on $\mathcal{D}$ may maximize energy absorption. This sensitivity analysis revealed the significance of each descriptor on different mechanical properties, providing a potential foundation for designing VACNT foams to achieve various independent property combinations tailored for a wide range of applications. 

To facilitate the implicit geometric descriptor-based design of architected VACNT foams for achieving tailored combinations of mechanical properties, we present the design contours for specific mechanical 
properties---$W^*$, $\sigma^*$, and $E^*$---predicted by our ANN model in Fig. \ref{fig6}(d-f). The design contour for $W^*$ (Fig. \ref{fig6}(d)) reveals a high-intensity region at lower values of $\mathcal{D}$, largely unaffected by variations in $\mathcal{I}_{mc-c}$. This aligns with the results of our sensitivity analysis, which indicated that $W^*$ can be effectively maximized by optimizing $\mathcal{D}$ alone. In contrast, the design contours for $\sigma^*$ (Fig. \ref{fig6}(e)) and $E^*$ (Fig. \ref{fig6}(f)) indicate that their maximum values require the simultaneous optimization of both $\mathcal{D}$ and $\mathcal{I}_{mc-c}$, reaffirming the insights from the sensitivity analysis. These design insights provide valuable guidance for developing architected VACNT foams tailored to specific applications. For example, materials with high energy absorption, strength, and modulus---characterized by large values of $W^*$, $E^*$, and $\sigma^*$---are ideal for protective systems such as shock absorbers \cite{gupta2024embracing}. Such optimal combination can be achieved through simultaneous optimization of the geometric descriptors $\mathcal{I}_{mc-c}$ and $\mathcal{D}$. On the other hand, functional applications like robotic arms---demanding materials with high energy dissipation for safe, impact absorbing interactions and high flexibility to adapt to irregular shapes \cite{cianchetti2018biomedical}---can be realized by designing foams with high $W^*$ and low $E^*$. This can be accomplished by minimizing $\mathcal{D}$ while maintaining higher values of $\mathcal{I}_{mc-c}$ as demonstrated by the design contours. By leveraging these design contours, one can systematically tailor the mechanical properties of architected VACNT foams to meet a wide range of applications, from robust, shock absorbing materials for protective systems to functional components with structural stability for soft robotics.

\begin{figure}[htp]
	\centering
	\includegraphics[width=160mm]{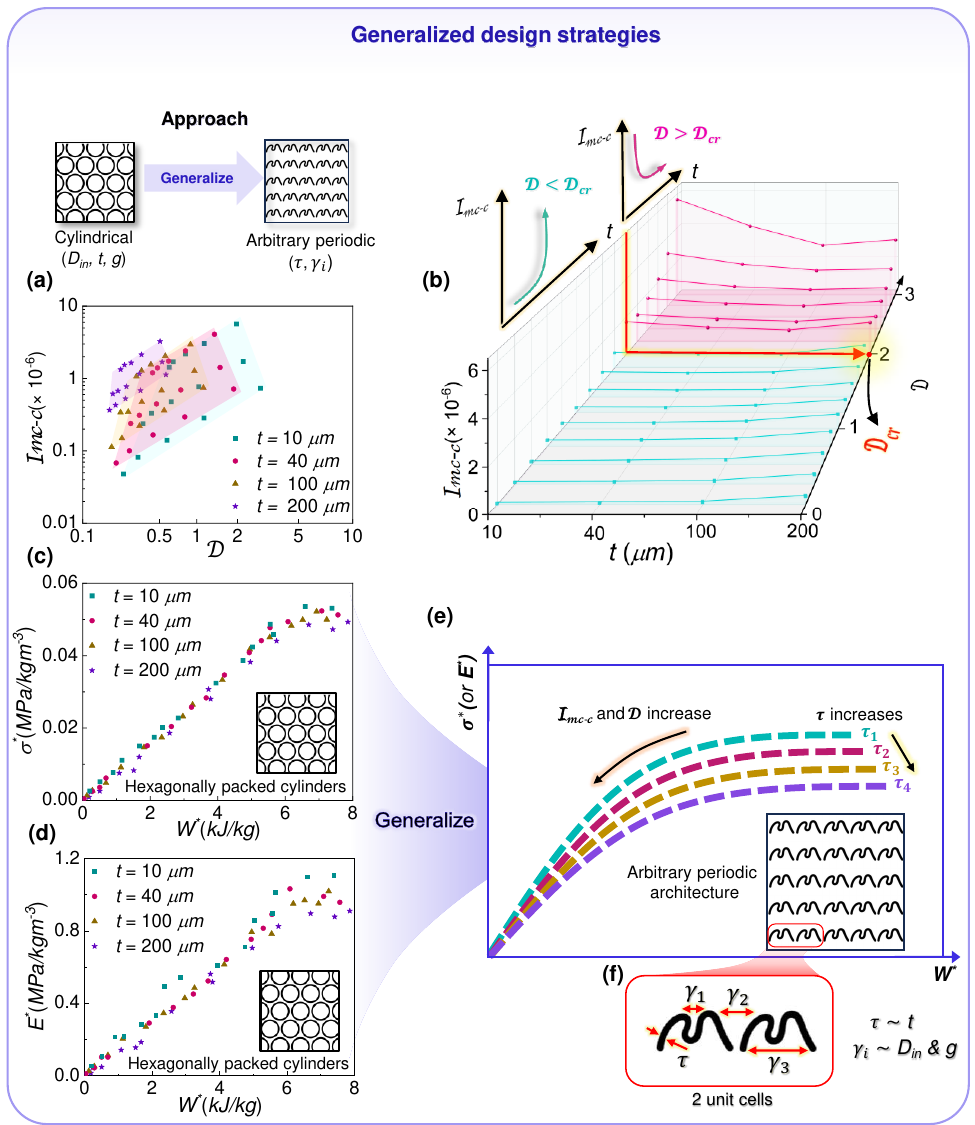}  
	\caption{Generalized design strategies. (a) Different constant thickness-trapezoidal regimes in the implicit descriptor landscape for cylindrical architectures enabling the decoupling between $\mathcal{I}_{mc-c}$ and $\mathcal{D}$ using $t$. (b) Relationship between $t$ and $\mathcal{I}_{mc-c}$ as a function of $\mathcal{D}$ revealing a reversal in the relationship at the critical point, $\mathcal{D}=\mathcal{D}_{cr}$ (When $\mathcal{D} < \mathcal{D}_{cr}$,  $\mathcal{I}_{mc-c}$ increases with $t$---enabling decoupling of descriptors; when  $\mathcal{D} > \mathcal{D}_{cr}$, $\mathcal{I}_{mc-c}$ decreases with $t$). Property landscapes predicted by the ANN model for cylindrical architectures revealing constant $t$ contours, along which descriptor values change: (c) $\sigma^*$ vs $W^*$ and (d) $E^*$ vs $W^*$. (e) A qualitative generalizable property landscape  $\sigma^*$(or $E^*$) vs. $W^*$, showing constant $\tau$ contours along which the descriptor values can be adjusted by fine-tuning $\gamma_i$ to navigate in the property landscape; Inset - an example of a periodic architectural design formed by expanding an arbitrary unit cell of constant thickness $\tau$ with (f) showing two arbitrary unit cells with constant thickness $\tau$ and other parameters that relate to the neighbor interactions $\gamma_i$. }
 	\label{fig7}
\end{figure}

\subsection{Generalizable design strategies for target mechanical properties}

While implicit geometric descriptor-based design offers a powerful approach to designing architected materials with diverse combinations of mechanical properties, its wide-spread adoption requires a clear understanding of pathways to independently control these descriptors. Ultimately, these descriptors should be manipulated by systematically adjusting the explicit geometric design parameters of a given architecture such as $D_{in}, t, g, g_{cc}, n$, and $m$ etc. To guide the control of implicit geometric descriptors---$\mathcal{I}_{mc-c}$ and $\mathcal{D}$---in unit-cell-based periodic architectural designs by the systematic adjustment of explicit design parameters, we utilized the set of 60 hexagonally packed cylindrical architectures as a model system to elucidate the trends between their explicit design parameters---$D_{in}, t$ and  $g$---and the implicit geometric descriptors.             

Fig. \ref{fig2}(a-c) illustrate the influence of explicit design parameters corresponding to cylindrical architectures on the descriptors, providing valuable insights into how these parameters can be adjusted to control the implicit descriptors in cylindrical architectures. Both $\mathcal{I}_{mc-c}$ and $\mathcal{D}$ exhibit monotonic trends with respect to $D_{in}$, $t$, and $g$, indicating predictable relationships. It is important to note that, the parameter $t$ has contrasting effects on the descriptors: $\mathcal{I}_{mc-c}$ increases with $t$, while $\mathcal{D}$ decreases  (Fig. \ref{fig2}(c)). This highlights $t$ as a potential decoupling design parameter, enabling independent manipulation of $\mathcal{I}_{mc-c}$ and $\mathcal{D}$. Fig. \ref{fig7}(a) shows the different trapezoid-shaped regimes (in the log scale) corresponding to different $t$ values that can be accessed in the geometric descriptor space of cylindrical architectures by selecting $t$ appropriately. This insight extends beyond cylindrical architectures to any periodic architecture formed by propagating a unit cell, as illustrated in Fig. \ref{fig7}(f), with its thickness, $\tau$---analogous to $t$ in cylindrical designs---and all other neighbor interaction governing parameters that define the empty spaces and gaps, $\gamma_1$,$\gamma_2$,...$\gamma_n$---analogous to $D_{in}$ and $g$ in cylindrical designs. After selecting an appropriate $\tau$ based on the desired range of $\mathcal{I}_{mc-c}$ and $\mathcal{D}$, the $\gamma_i$ can be fine-tuned to precisely adjust the descriptor values. This approach simplifies pathways to achieve desired combinations of geometric descriptor values, which, as discussed earlier, are crucial for designing architectures with diverse mechanical property combinations.     

Although $\mathcal{I}_{mc-c}$ generally increases with $\tau$, it may decrease with $\tau$ for periodic architectures having large $\gamma_i$ values. As illustrated in Fig. \ref{fig1}(d), $\mathcal{I}_{mc-c}$ was shown to decrease with both the size of the unit cell and its population. However, an increase in unit cell size inevitably leads to a decrease in unit cell population due to spatial constraints. Consequently, a competing effect emerges between these two factors, leading to a trend reversal in the $\mathcal{I}_{mc-c}$ vs. $\tau$ relationship at a critical point within the $\mathcal{D}$ space as $\gamma_i$ increase. Fig. \ref{fig7}(b), shows how the relationship between $t$ (analogous to $\tau$) and $\mathcal{I}_{mc-c}$ evolves with $\mathcal{D}$ in cylindrical architectures where the trend reversal is clearly demonstrated. At a critical value, $\mathcal{D} = \mathcal{D}_{cr}$, the relationship shifts: in the region  where $\mathcal{D} < \mathcal{D}_{cr}$ (cyan), $\mathcal{I}_{mc-c}$ increases with $\tau$, while it decreases in the region where $\mathcal{D} > \mathcal{D}_{cr}$ (pink) (Fig. \ref{fig7}(b)). Therefore, to effectively utilize the design insights discussed previously, this critical point must be identified and characterized for each specific unit cell type. Thus, the design region preceding this critical point ($\mathcal{D} < \mathcal{D}_{cr}$) can then be exploited to decouple $\mathcal{I}_{mc-c}$ and $\mathcal{D}$, enabling tailored control over mechanical properties. 

Understanding how tuning $\tau$ and $\gamma_i$ maps onto the mechanical property landscape further refines our strategy. Although $\tau$ may influence $\mathcal{I}_{mc-c}$ differently depending on the region in the $\mathcal{D}$ space, it is intuitive that increasing $\tau$ generally reduces $\sigma^*$ and $E^*$ owing to the size effect and the reduction in the unit cell population as the unit cell size increases. On the other hand, increasing $\tau$ decreases $\mathcal{D}$ (Fig. \ref{fig1}(d))---the predominant influencer of $W^*$---resulting in enhancement of $W^*$ (Fig. \ref{fig6}(d)). These trends collectively suggest that by increasing $\tau$, periodic architectures with low $\sigma^*$ and $E^*$ can be achieved while maintaining high $W^*$. Fig. \ref{fig7}(c) and Fig. \ref{fig7}(d) respectively illustrate the $\sigma^*$ vs. $W^*$ and $E^*$ vs. $W^*$ property landscapes for cylindrical architectures, highlighting constant $t$ (analogous to $\tau$) contours as predicted by the ANN model. As $t$ increases, the contours shift towards right and stay below the smaller $t$ contours indicating higher $W^*$ and lower $\sigma^*$ and $E^*$. However, along each contour, all three mechanical properties simultaneously increase in the direction where $\mathcal{I}_{mc-c}$ and $\mathcal{D}$ simultaneously decrease due to variations in parameters such as $D_{in}$ and $g$ (analogous to $\gamma_i$).
Guided by these behaviors, we propose a generalizable, qualitative $\sigma^*$(or $E^*$) vs. $W^*$ property landscape (Fig. \ref{fig7}(e)) for an arbitrary unit cell-based periodic architecture (Fig. \ref{fig7}(e)(Inset)). This landscape presents qualitative constant $\tau$ contours capturing the insights gained from the cylindrical architectures. With the selection of an appropriate $\tau$ to define a target property regime, the $\mathcal{I}_{mc-c}$ and $\mathcal{D}$ can be adjusted by fine-tuning $\gamma_i$ to navigate along the contours to realize desired combination of mechanical properties. By leveraging this implicit geometric descriptor enabled-design framework one can simplify the design process and realize systematic pathways for tailoring the mechanical performance of architected VACNT foams across a broad spectrum of applications.

\section{Conclusion}

We introduced a design framework to establish a unified structure-property relationship in architected VACNT foams by leveraging  MCSI implicit geometric descriptors which capture the geometric features of the architecture that reflect nanoscale mechanisms---such as size effect and inter-nanotube interaction---that predominantly govern the bulk specific mechanical properties of the architected VACNT foams. By correlating these implicit descriptors to the experimentally measured mechanical properties using an ANN framework, we establish a robust and generalizable structure-property relationship that holds across diverse architectures. We also demonstrated the strong predictive capability of this approach by accurately estimating the mechanical properties for entirely new unseen designs, including non-cylinder-based architectures such as sinusoid-based designs despite the model being trained exclusively on cylinder-based designs. This remarkable generalization power can enable a wide range of architectural designs. Importantly, our approach was able to establish this cohesive structure-property relationship using just two key geometric descriptors that encapsulate both size and interaction effects. This suggests that despite the apparent influence of the explicit geometric parameters of a particular architecture, the macroscopic mechanical behavior of this material system is largely governed by the nanoscale mechanisms such as the size effects and the complex interactions among nanotubes.

Our study not only advances the understanding of the foundational mechanisms governing the mechanical properties of architected VACNT foams but also offers a framework that can be extended to other architected nanofibrous and micro-architected material systems where similar nanoscale mechanisms interactively influence macroscopic mechanical properties. A key innovation of this study lies in the use of implicit geometric descriptors with meaningful physical relevance within an ANN framework, rather than relying on image regression-based and other parameter-heavy deep learning approaches. This shift has the potential not only to accelerate and streamline the design of architected materials but also to enhance the interpretability and tractability of the design process.

\begin{figure}[htp]
\textbf{Graphical Abstract}
\centering
\includegraphics[width=\textwidth]{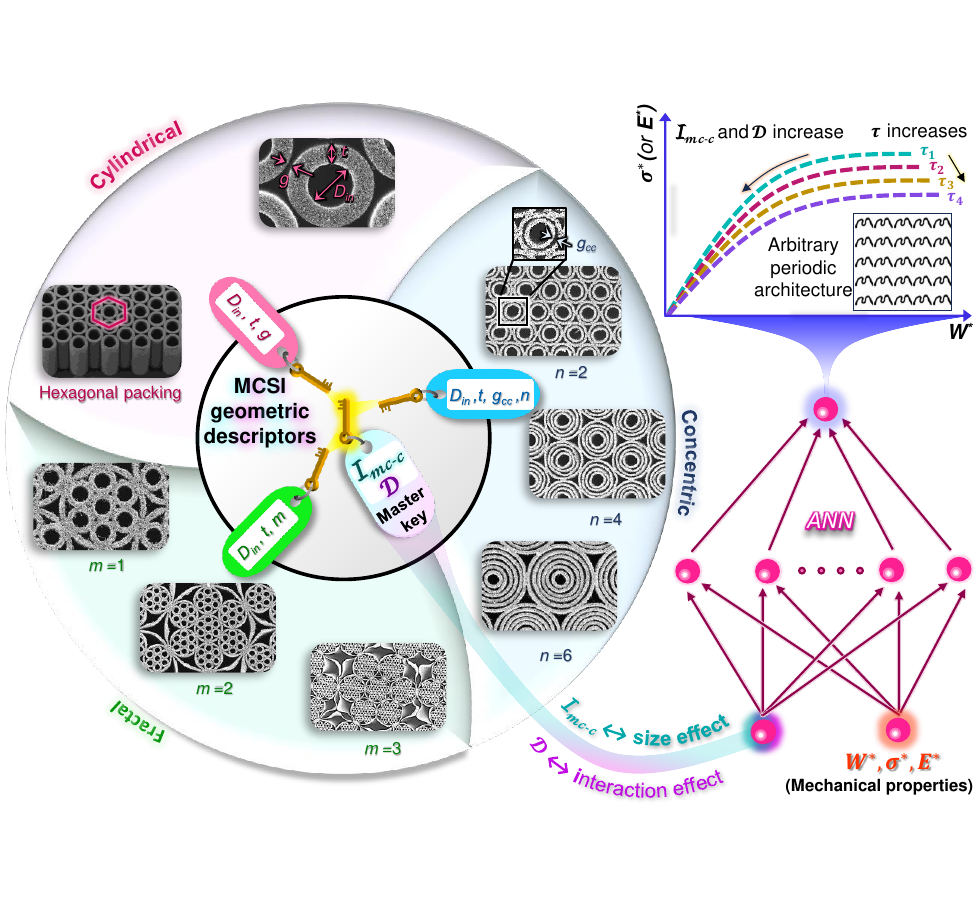}  
\caption{\textbf{Graphical Abstract:} MCSI implicit geometric descriptor-enabled ANN framework establishes a unified structure property relationship across diverse architectures and enables a generalized foundation for the accelerated and tractable design of architected nanofibrous materials for target mechanical properties}
 	\label{figGA}
\end{figure}

\newpage
\section*{Acknowledgments}

 The authors acknowledge the support from the U.S. Office of Naval Research under the PANTHER program award numbers N00014-21-1-2044 and N00014-24-1-2200 through Dr. Timothy Bentley. This work also utilized facilities and instrumentation at the Wisconsin Centers for Nanoscale Technology (WCNT), which is partially supported by the National Science Foundation through the University of Wisconsin Materials Research Science and Engineering Center (DMR-1720415). We extend our thanks to Professor Chris Rycroft for his valuable insights on describing geometries within the architecture, and to Daniyar Syrlybayev for his assistance with some of the scanning electron microscopy.
 \section*{Author contributions}

\textbf{Bhanugoban Maheswaran}: Methodology, Investigation, Validation, Formal analysis, Writing-original draft. \textbf{Komal Chawla}: Methodology, Investigation, Data curation. \textbf{Abhishek Gupta}: Investigation, Data curation. \textbf{Ramathasan Thevamaran}: Conceptualization, Methodology, Writing – review \& editing, Supervision, Project administration, Funding acquisition.


\section*{Resource availability}
\subsection*{Lead contact}

Requests for further information and resources should be directed to and will be fulfilled by the lead contact, Ramathasan Thevamaran (thevamaran@wisc.edu).

\subsection*{Materials availability}

The data used to support the findings of this study are available in Supplementary Information

\subsection*{Code availability}

MATLAB codes for the generating patterns, computing MCSIs,  data processing and ANN training \color{black}are available on \url{https://github.com/ThevamaranLab/MCSI-enabled-ANN-framefwork}

\section*{Supplementary Information}

\begin{description}

\item Figure S1 available in Supplementary information (PDF)
\item Figure S2 available in Supplementary information (PDF)
\item Figure S3 available in Supplementary information (PDF)
 \item Table S1, S2 , S3, S4 and S5 available in Supplemental information (PDF)

   \end{description}

\section*{Competing Interests}

The authors declare no competing interest.

\label{section:sd}

\bibliographystyle{unsrtnat}
\bibliography{Manuscript_EML_BM_KC_AG_RT.bib}

\end{document}